\documentclass[a4paper]{jpconf}
\usepackage{citesort}
\usepackage[dvipdfmx]{graphicx}
\usepackage{amsmath,amssymb}
\usepackage{bm}
\begin{document}
\title{Quadrupole order in the frustrated pyrochlore magnet Tb$_2$Ti$_2$O$_7$}

\author{H. Takatsu$^{1,2}$, T. Taniguchi$^1$, S. Kittaka$^3$, T. Sakakibara$^3$, and H. Kadowaki$^1$}
\address{$^1$Department of Physics, Tokyo Metropolitan University, Hachioji-shi, Tokyo 192-0397, Japan}
\address{$^2$Department of Energy and Hydrocarbon Chemistry, Graduate School of Engineering, Kyoto University, Kyoto 615-8510, Japan}
\address{$^3$Institute for Solid State Physics, University of Tokyo, Kashiwa 277-8581, Japan}

\begin{abstract}
We have studied the hidden long-range order (LRO) of the frustrated pyrochlore magnet Tb$_2$Ti$_2$O$_7$
by means of specific-heat experiments and Monte-Carlo (MC) simulations,
which has been discussed as the LRO of quadrupole moments
inherent to the non-Kramers ion of Tb$^{3+}$.
We have found that the sharp specific-heat peak is collapsed into a broad hump
by magnetic fields above 0.3 T for $H//[001]$.
This result, qualitatively reproduced by MC simulations, suggests that 
a field-induced magnetic state overcomes
the quadrupolar LRO state, as a similar case of a classical spin ice. 
The present results support the interpretation that Tb$_{2+x}$Ti$_{2-x}$O$_{7+y}$ is a unique material 
in the boundary between the quadrupolar ($x \geq x_c = -0.0025$) and spin-liquid ($x \leq x_c$) states,
where the magnetic field along the [001] axis is a tuning parameter which induces the magnetic ordered state.
\end{abstract}

\section{Introduction}
Geometrically frustrated magnetic systems provide us with a rich playground for studying new type of electronic and 
magnetic behaviors with unconventional order parameters~\cite{C.Lacroix,GingrasRPP2014}. 
In particular, the pyrochlore-lattice magnet Tb$_{2+x}$Ti$_{2-x}$O$_{7+y}$,
a putative candidate of spin liquid (SL)~\cite{GardnerPRL1999,GardnerPRB2003},
is a unique material showing an unknown long range order (LRO) in the vicinity ($x \geq x_c = -0.0025$) of the SL state~\cite{TaniguchiPRB2013,Wakita_singl_crystal_TTO}.
It is found that a clear specific-heat peak appears at $T_{\rm c} \simeq 0.5$~K for the sample with $x = 0.005$, 
while no LRO associated with the large magnetic and/or structural phase transitions was observed~\cite{TaniguchiPRB2013}:
the only small Bragg peak (the order of 0.1~$\mu_{\rm B}$/Tb) appears below $T_{\rm c}$
but is too small to explain the corresponding entropy change in the specific heat.
Therefore,
the LRO of Tb$_2$Ti$_2$O$_7$ is considered to be a ``hidden order''~\cite{SantiniRMP2009}, 
apparently contrast with the magnetic dipole order inferred by earlier theories 
on the basis of the spin ice (SI) Hamiltonian~\cite{GingrasPRB2000,KaoPRB2003}. 

Recently, we have investigated the single crystalline sample of Tb$_{2+x}$Ti$_{2-x}$O$_{7+y}$ with $x = 0.005$ ($T_{\rm c} = 0.53$~K),
and found experimental key ingredients of the hidden order~\cite{Takatsu_TTO2}; 
i.e., two-step magnetization kink and 
double peak structure in specific heat, which are induced by magnetic field for $H//[111]$.
We have demonstrated those as characteristics of the behaviors of a possible electric quadrupole ordering~\cite{Takatsu_TTO2},
by using classical Monte Carlo (CMC) and quantum Monte Carlo simulations on the basis of the theoretical model
proposed by Onoda and Tanaka~\cite{S.Onoda2010PRL,S.Onoda2011PRB}. 
This theory shows that transverse super-exchange interactions between quadrupole moments, 
which are set into the SI Hamiltonian as an additional term, 
play a driving force of the LRO, making Tb$_2$Ti$_2$O$_7$ into a type of a quadrupole order 
in the vicinity of the SL state~\cite{S.Onoda2010PRL,S.Onoda2011PRB,LeePRB2012}.
Following this interpretation,
we have also pointed out that the appearance of the inelastic neutron excitation 
can be understood in terms of the spin-quadrupole wave~\cite{H.KadowakiSPIN2015,Takatsu_TTO2}. 
These results suggest that the problem of the hidden order in Tb$_2$Ti$_2$O$_7$ can now be reconsidered as the novel problem of 
the quadrupole order in the frustrated pyrochlore magnet with the non-Kramers ion of Tb$^{3+}$. 
%

The purpose of the present study is to examine the quadrupole order of Tb$_2$Ti$_2$O$_7$ 
by measuring physical quantities in magnetic field along other directions for $H//[111]$, 
which provide another evidence for the origin of the hidden order
and may induce exotic magnetic properties. 
For this purpose, we performed specific heat experiments under magnetic field for $H//[001]$. 
CMC simulations are also performed. 
We found that the sharp peak of the specific heat at $H = 0$ is suddenly suppressed by the magnetic field along [001].
This behavior is understood in terms of the realization of 
the field-induced magnetic state in weak [001] fields above 0.3~T.

\section{Experimental}
Single crystals of Tb$_{2+x}$Ti$_{2-x}$O$_{7+y}$ were grown by a floating zone method~\cite{Wakita_singl_crystal_TTO}.
We used the crystal with $x = 0.005$ in this study.
Specific heat $C_P(T,H)$ was measured by a quasi adiabatic method in a dilution refrigerator down to 0.1~K.
Magnetic field was applied using a vector magnet system where
an accuracy of the field direction to the sample is below $1^\circ$.
In order to reduce the demagnetization effect,
we used a plate-like crystal, cut out from a single crystal rod, 
along the $\langle110\rangle$ plane which includes the [111], [110], and [001] axes.
The sample is approximately $0.7\times0.9\times0.1$~mm$^3$ and weighs 0.35~mg: 
this crystal is the same crystal used in the previous study~\cite{Takatsu_TTO2}.
Since the demagnetization factor for the [001] direction is small enough ($N\simeq0.09$),
demagnetization corrections were not performed in the present study.

CMC simulations were performed using the program of the ALPS package 
(1024 sites, periodic boundary conditions)
on the basis of the effective pseudospin-1/2 Hamiltonian~\cite{S.Onoda2010PRL,S.Onoda2011PRB} relevant
for the non-Kramers magnetic doublets of Tb$_2$Ti$_2$O$_7$.
We employed the nearest-neighbor (NN) classical SI model for the first three terms of Eq.~(1) of Ref.~\cite{Takatsu_TTO2}
and used following parameters, $J_{\rm nn, eff} = 1.48$~K, $J_{\rm nn}*\delta = 0$,  and $J_{\rm nn}*q = 0.85$, 
so that those correspond to the previously used parameters~\cite{Takatsu_TTO2}.
These parameters are in the region of the planar antiferropseudospin (PAF) state.
Readers are referred to Ref.~\cite{Takatsu_TTO2} for details of the Hamiltonian and the parameters.
We calculated the specific heat under the [001] field in the present study.
We also computed the order parameter of the quadrupolar LRO,
which is a type of magnetization associated with the LRO of the $xy$-components of the Pauli matrices $\bm{\sigma}_{ \bm{r} }$:
the quadrupolar LRO can be expressed by a pseudospin structure $(\langle \sigma_{ \bm{r} }^x \rangle , \langle \sigma_{ \bm{r} }^y \rangle )$
in the model of Refs.~\cite{S.Onoda2010PRL,S.Onoda2011PRB}.

\section{Results and Discussion}
\begin{figure}
\begin{center}
\includegraphics[width=0.80\textwidth,clip]{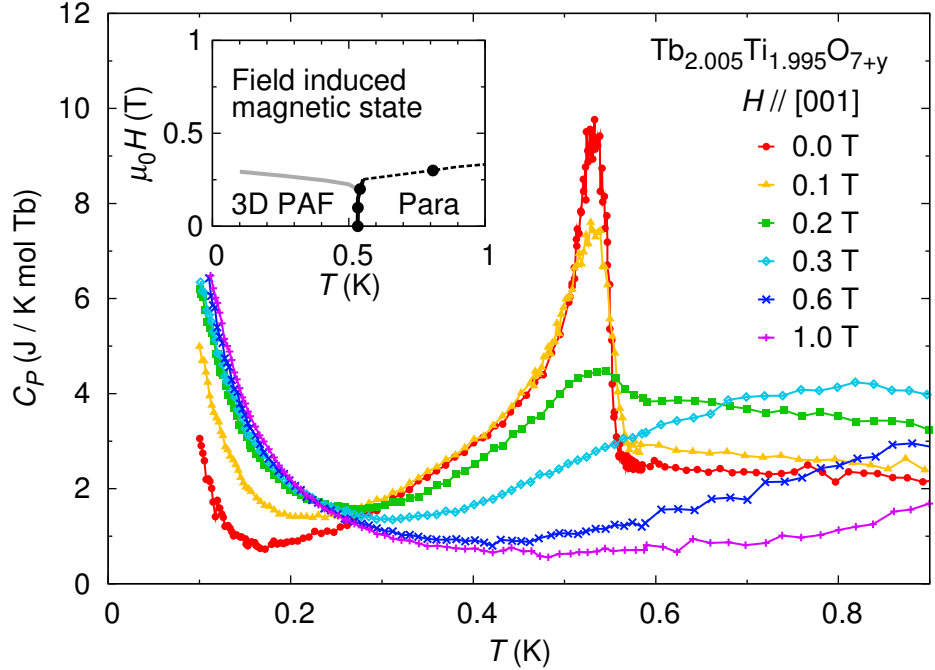}
\end{center}
\caption{
Temperature dependence of the specific heat $C_P(T,H)$ of single-crystalline
Tb$_{2+x}$Ti$_{2-x}$O$_{7+y}$ with $x=0.005$ in the applied magnetic field along [001].
Inset shows the $H$--$T$ phase diagram for $H // [001]$.
Filled circles in the inset are peak positions of $C_P(T,H)$.
The black solid line up to 0.2~T indicates the phase boundary between the quadrupolar (3D PAF) state 
and the paramagnetic state. The black dashed line is the line of the crossover between the paramagnetic state 
and the field-induced magnetic state.
The gray solid line is the putative phase boundary between the 3D PAF state and the field-induced magnetic state,
which is examined from the analysis of CMC simulations. 
Labels in the phase diagram are assigned from the analysis of CMC simulations (Fig.~\ref{fig.3}).
}
\label{fig.1}
\end{figure}
\begin{figure}
\begin{center}
\includegraphics[width=0.80\textwidth,clip]{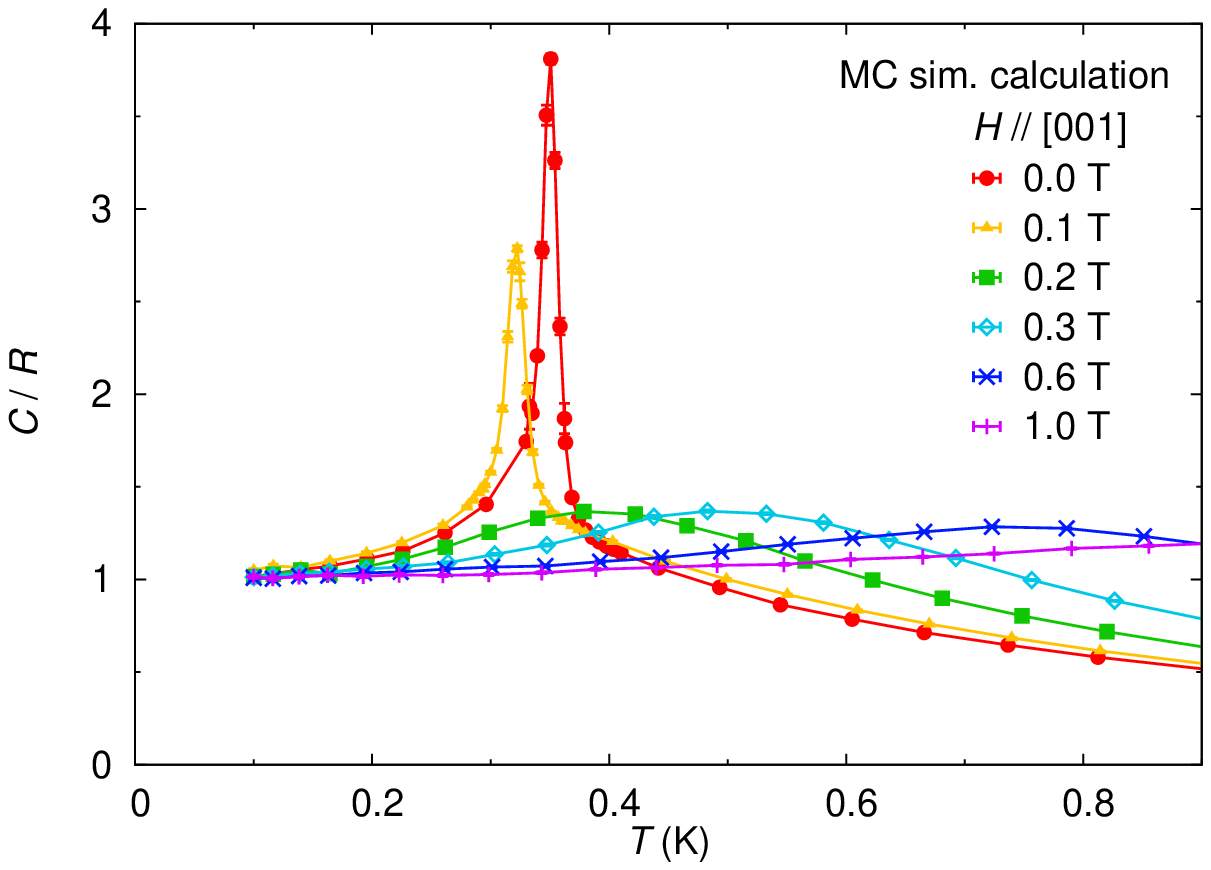}
\end{center}
\caption{
Temperature dependence of the specific heat for several fields
calculated by the CMC simulation. 
}
\label{fig.2}
\end{figure}
Figure~\ref{fig.1} shows the temperature dependence of $C_P(T,H)$
measured at several magnetic fields up to 1~T.
The sample exhibits a clear sharp peak at $T_{\rm c} = 0.53$~K in zero field,
which is compatible with the behavior of the polycrystalline sample of $x = 0.005$~\cite{TaniguchiPRB2013}.
Remarkably, once a magnetic field is applied along [001],
the sharp peak is collapsed into a broad hump in fields above 0.3~T.
This result suggests that 
the zero-field LRO state is weak for the application of magnetic field for $H//[001]$.
In the inset of Fig.~\ref{fig.1},
we summarized the $H$--$T$ phase diagram for $H//[001]$.
The labels of each phase were determined from the analysis of CMC simulations, which will be discussed later.
We found that the quadrupole LRO is replaced by the field-induced magnetic ordered state for $H//[001]$.
This result is related to the crystal structure of the pyrochlore lattice and 
the Ising-like anisotropy of spins along the local $\langle 111 \rangle$ direction. 
In fact, for magnetic field along the [001] axis,
it is expected that magnetic moments are easily induced 
as those pointing at the inward and outward tetrahedra consisting of the 2-in 2-out spin configurations. 
It is known for a classical SI material for $H//[001]$~\cite{FukazawaPRB2002,HiroiJPSJ2003}.

Figure~\ref{fig.2} shows the temperature dependence of the calculated specific heat.
We confirmed that
the experimental behaviors are qualitatively reproduced by CMC simulations, 
although the slight lowering of $T_{\rm c}$ is observed in low field.
In fact, we observed the abrupt change in the peak height of the specific heat 
in CMC simulations for $H//[001]$.
This result implies that the present model~\cite{S.Onoda2010PRL,S.Onoda2011PRB} 
works well for the discussion for the LRO of Tb$_2$Ti$_2$O$_7$,
where the dominant component for the formation of LRO is the super-exchange interaction between quadrupole moments.
%

\begin{figure}
\begin{center}
\includegraphics[width=0.95\textwidth,clip]{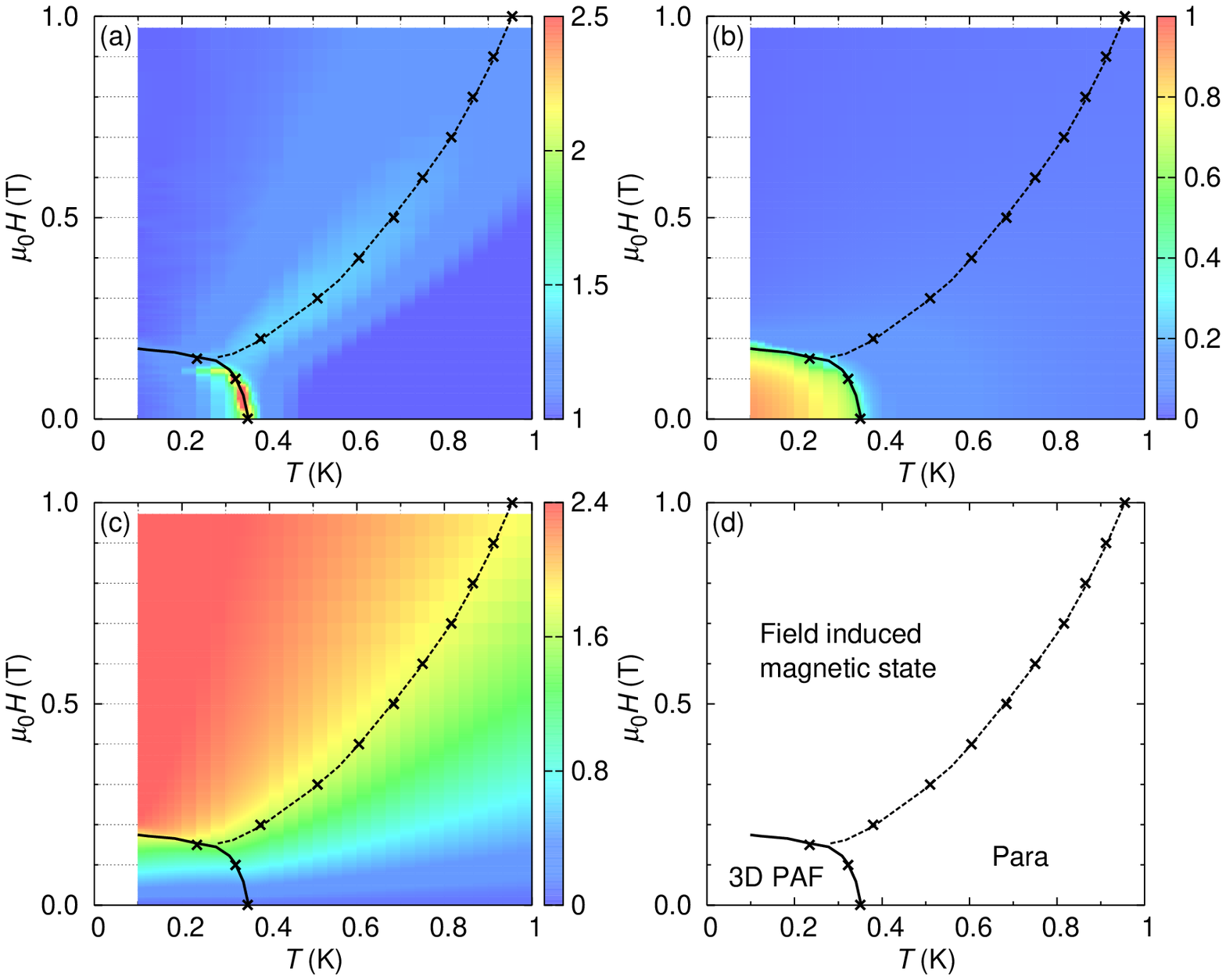}
\end{center}
\caption{
Temperature-field dependence of (a) specific heat, (b) quadrupolar order parameter, and (c) magnetization
calculated by the CMC simulation for magnetic field along [001].
(d) $H$--$T$ phase diagram obtained by the CMC simulation for $H // [001]$.
Cross symbols in (a)--(d) are peak positions of the calculated specific heat.
The solid lines indicate the phase boundary between the 3D PAF state and the field-induced magnetic state or the paramagnetic state.
The dashed lines are the lines of the crossover between the paramagnetic state and the field-induced magnetic state.
The lines are guides to the eyes.
}
\label{fig.3}
\end{figure}
In Figs.~\ref{fig.3}(a)--(c),
we present contour plots of the calculated specific heat,
order parameter of the quadrupolar LRO, and magnetization $M$.
%
The map of specific heat shows the qualitatively similar behavior for experiments,
where the abrupt change appears around $T_{\rm c}$ in low fields and the broad hump shifts to high temperatures
when the field is increased.
The phase boundary extending to low temperatures is less obvious,
however it can be expected by the appearance and disappearance of order parameters as shown in Figs.~\ref{fig.3}(b) and (c).
In fact, 
we confirmed that the order parameter of quadrupole moments develops in the low-$H$ and low-$T$ region below $T_{\rm c}$.
Here, this quadrupolar state is found to be the three dimensional (3D) PAF state~\cite{H.KadowakiSPIN2015,Takatsu_TTO2}.
However, it is soon destroyed by the large magnetic field for $H//[001]$ (Fig.~\ref{fig.3}(b)).
We realized that $M$ is developed in such field region (Fig.~\ref{fig.3}(c)).
The value of $M$ in the phase boundary is about 70\% of the full magnetic moment for the [001] direction ($\sim2$~$\mu_{\rm B}$/Tb):
note that 
the present CMC simulations can only handle the contribution of the magnetic moment of 
the ground state doublet ($\mu\simeq5$~$\mu_{\rm B}$/Tb)~\cite{H.KadowakiSPIN2015,Takatsu_TTO2},
and then the fully saturated moment of $M$ along the [001] direction is $\mu/\sqrt{3}$ due to 
the anisotropy of the [001] direction~\cite{FukazawaPRB2002}.
Interestingly, we found that the boundary, accompanied by the specific-heat peak and the disappearance of the quadrupole order parameter, 
emerges at the line of such 70\% magnetization (Figs.~\ref{fig.3}(a)--(c)). 
This result implies a phase boundary between the quadrupolar LRO state and the field-induced magnetic state or the paramagnetic state ($T<T_{\rm c}$),
and a crossover from the paramagnetic state to the field-induced magnetic state ($T>T_{\rm c}$) (Fig.~\ref{fig.3}(d)).
This situation is probably realized in the experiments for $H//[001]$.
In fact, in the previous magnetization experiment by Legl \textit{et al}.~\cite{LeglPRL2012},
the sharp increase of $M$ has been observed for $H//[001]$.
The field where the magnetization reaches 2~$\mu_{\rm B}$/Tb is 0.3 -- 0.4~T at $T\leq0.3$~K. 
This is compatible with the present CMC calculations and the experimental results.

Studies of bulk physical properties in magnetic field are indirect observation for the origin of the quadrupole ordering.
However, we can observe key signatures associated with the formation and deformation of its ordering.
Therefore, the present specific heat experiment for $H//[001]$ and analysis by CMC simulations 
are important clues to discuss the origin of the LRO of Tb$_2$Ti$_2$O$_7$,
which should be a quadrupole ordering and is weak for the [001] field .
We can now consider that Tb$_2$Ti$_2$O$_7$ is a very unique material showing two ground states, the quadrupolar-LRO state and SL state,
which are tuned by the minute change in the off-stoichiometric parameter of $x$ for Tb$_{2+x}$Ti$_{2-x}$O$_{7+y}$~\cite{TaniguchiPRB2013,Wakita_singl_crystal_TTO}.
Magnetic field is also a tuning parameter which instead induces the magnetic ordered state for $H//[001]$.

\section{Conclusion}
In conclusion,
we have studied the long-range order (LRO) of the frustrated pyrochlore-lattice magnet Tb$_2$Ti$_2$O$_7$,
which emerges in the vicinity of the spin liquid state by tuning the parameter $x$ for Tb$_{2+x}$Ti$_{2-x}$O$_{7+y}$.
We measured the specific heat of the single crystal with $x=0.005$ ($T_{\rm c} = 0.53$~K) and 
found that the sharp specific-heat peak is suddenly suppressed by magnetic field for $H//[001]$.
We demonstrated that the experimental behavior is qualitatively reproduced by the classical Monte Carlo simulation
on the basis of the Onoda-Tanaka Hamiltonian.
These results imply that the field-induced magnetic ordered state overcomes the quadrupolar LRO state at magnetic fields above 0.3~T.
This behavior is characteristic for the magnetic field along [001].
The present experimental and calculation results support the scenario that 
the LRO of Tb$_2$Ti$_2$O$_7$ is a quadrupole order, mediated by a quantum phase transition from a spin liquid state,
where the magnetic field can be used as a tuning parameter as well for the realization of a magnetic ordered state.
%

\ack 
We thank S. Onoda, Y. Kato, R. Higashinaka and M. Wakita for useful discussions. 
This work was supported by JSPS KAKENHI grant numbers 25400345 and 26400336.
The specific heat measurement was performed using the facilities of ISSP, Univ. of Tokyo.
\section*{References}
\bibliography{TTO_HT.bib}

\end{document}